# Report by the Committee on the Scientific Case of the ILC Operating at 250 GeV as a Higgs Factory

**July 22, 2017**

Committee Members:
Shoji Asai[1,2,*], Junichi Tanaka[2], Yutaka Ushiroda[3], Mikihiko Nakao[3], Junping Tian[2], Shinya Kanemura[4], Shigeki Matsumoto[5], Satoshi Shirai[5], Motoi Endo[3], Mitsuru Kakizaki[6]

* Chair
[1] The University of Tokyo
[2] ICEPP, The University of Tokyo
[3] High Energy Accelerator Research Organization (KEK)
[4] Osaka University
[5] Kavli IPMU, The University of Tokyo
[6] University of Toyama

Commissioned by the Japan Association of High Energy Physicists

## Preface

In July 2012, a Higgs boson with 125 GeV mass was discovered at the LHC. The discovery of new phenomena and new principles that can (naturally) explain the electroweak symmetry breaking (EWSB) including the existence of this Higgs boson is now the most important and urgent target of research. In order to attain this goal, the LHC is performing direct searches for new phenomena and new principles with a center-of-mass (CM) energy increased to 13 TeV. So far, there is no evidence of new physics beyond the Standard Model (SM). The purpose of this committee is to investigate and compare, under the current circumstances, the capability to determine the energy scale of new phenomena and new principles and the capability to uncover the origin of matter-antimatter asymmetry for the following three cases: (i) an ILC operating at 250 GeV as a "Higgs Factory" (ILC250); (ii) an ILC operating up to 500 GeV (ILC500); and (iii) the case of no ILC construction. The committee members consist primarily of members of the ATLAS collaboration, the Belle II collaboration, and theorists. The committee aimed to give an assessment on the physics case of the ILC250 in a way that is independent from the ILC community.

This report consists of the following five chapters:

1. Introduction
2. Precise measurements of Higgs and other SM processes: Determination of the energy scale of new phenomena via precise measurements.
3. EWSB and the origin of matter-antimatter asymmetry.
4. Direct search for dark matter and new particles based on "Naturalness".
5. Summary: Comparison of the ILC operating at 250 GeV and 500 GeV

Different approaches are summarized in Chapters 2 and 3 to probe the next energy scale beyond EWSB through precise measurements. Chapter 4 discusses searches to elucidate dark matter (DM) and probes to test the idea of "naturalness".

## 1. Introduction

For the purpose of this discussion, the following points are assumed for the timeline and the conditions of the ILC operation.

1. The operation will start around 2028-2030. It will run concurrently with the High-Luminosity LHC (HL-LHC) experiment and produce complementary results.
2. The CM energy is fixed at 250 GeV. No energy scan is performed. The integrated luminosity is 200 $fb^{-1}$ per year, accumulating 2 $ab^{-1}$ by 2040.

3. Beam polarization is used (30% for positrons, 80% for electrons).

The key is the synergy with other experiments, including the HL-LHC, SuperKEKB, Hyper-Kamiokande, electric dipole moment (EDM) searches, lepton flavor violation (LFV) searches, and satellite probes to detect gravitational waves (LISA, DECIGO, etc.), as well as theory development in Lattice QCD and higher-order corrections. Various implications are considered combining rich outputs from these experiments with the ILC results, and we elucidate the role of the ILC with respect to the other experiments.

## 2. Higgs and Other Standard Model Processes: Determination of the New Energy Scale via Precision Measurements

### 2.1. Precise measurements of Higgs couplings

The precise measurements of the couplings between Higgs boson and other elementary particles can be performed at the 0.6-1.8% level at the ILC250. The measurement precisions are summarized in Tables 1 and 2. It is the important task of ILC to measure the total decay width model-independently. The previous strategy is as follows, the HWW coupling is measured using the vector boson fusion process, and decay branching fraction Br(H→WW) can be measured precisely. Then total decay can be determined model-independently. The vector boson fusion process enables the precise measurement of the HWW coupling at higher energies. At the ILC250, however, this cross section is small. It was one of motivations for higher center of mass at ILC.

As an alternative approach, we can measure Higgs decay branching fraction at the HL-LHC to examine the symmetry between the HWW and HZZ couplings (custodial symmetry) at the 2% level. By taking this symmetry as an assumption, the ee→ZH cross section (HZZ coupling) and the H→WW decay branching ratio measurements can be combined for the model-independent determination of the total decay width. This idea can be further extended in the framework of effective field theories to determine the coupling (denoted as g), in a model-independent way (Ref. airXiv 1708.09079).

The estimated precisions of various Higgs boson couplings are shown in Table 1, combining ILC250 and HL-LHC results. The precisions are at the 10% level with the HL-LHC alone, less than 1% accuracies can be obtained as shown in Table 1. The comparison between ILC250 and ILC500 shown in Fig.1 shows that the differences in the achievable precisions are small for the same total integrated luminosity of 2 $ab^{-1}$. This illustrates the importance of the combination of the HL-LHC and the ILC250 results. These combined results are comparable to those at ILC500 (combined with HL-LHC). Table 2 summarizes the precision of the coupling ratios from the direct determination at ILC250. Many experimental systematic uncertainties cancel by taking the ratios. These ratios are useful for the precise comparison between the SM predictions and the experimentally observed values.

Table 1: Precision of Higgs boson couplings in the effective field theory framework. Combination of the ILC250 and the HL-LHC measurements

| | g(HZZ) | g(HWW) | g(Hbb) | g(Hττ) | g(Htt) | g(Hμμ) | g(Hcc) |
|---|---|---|---|---|---|---|---|
| Δg/g | 0.63% | 0.63% | 0.89% | 1.0% | 7% (LHC) | 6.2% | 1.8% |

Table 2: Precision of Higgs coupling ratios from the direct measurements at the ILC250.

| | g(HWW)/g(HZZ) | g(Hbb)/g(HWW) | g(Hττ) /g(HWW) | g(Hcc) /g(HWW) |
|---|---|---|---|---|
| Δ | 1.9% | 0.64% | 0.84% | 1.7% |

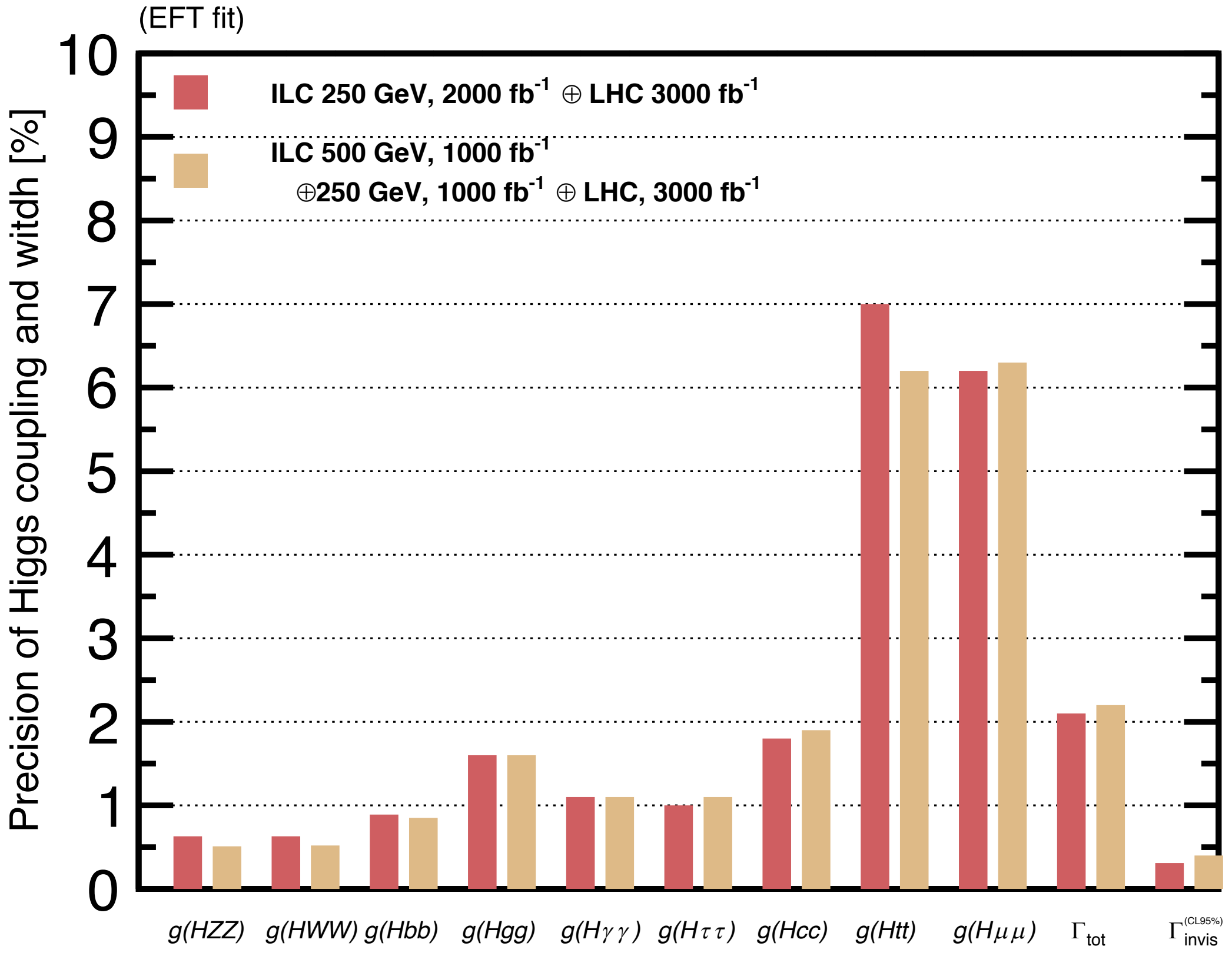

Figure 1: Precision of coupling measurements.

It will be possible to measure the Yukawa couplings of the second-generation leptons and quarks with about 2-6% precision. This will show that the differences in the Higgs boson couplings give rise to the generations, providing insight into our understanding of generations. Furthermore, the measurement of the Higgs boson coupling with the Z boson and the differential cross section can be done model-independently and provides sensitivity to new phenomena with a mass scale of up to 2.5 TeV for CP-even states and 3.9 TeV for CP-odd states using the effective Lagrangian approach.

Among the various new phenomena and new principles, the supersymmetry is the most promising theory. There are three approaches to discover supersymmetric particles, as described below. The capability of these three approaches depends on the model and the parameter space. Thus it is crucial to be able to cover all three approaches.

(a) Direct search for supersymmetric partners with SU(3) color charge, such as the squarks and the gluino. The HL-LHC has good sensitivities to search for the squarks and gluino up to about 3 TeV.

(b) Search for supersymmetric partners with SU(2) and U(1) charge, such as the electroweak gauginos and higgsinos. In contrast to (a), the mass spectrum can be naturally highly compressed. The ILC will play an important role as described in Chapter 4.

(c) There are at least two Higgs doublets (2HD) in all supersymmetric models, in which multiple Higgs bosons exist. These signatures can be accessed even if the supersymmetric partners as described in (a) and (b) are beyond the reach of the experiments.

The precise measurement of the Higgs couplings at the ILC250 provides an important input to the approach (c) above. The Minimal Supersymmetric Standard Model (MSSM) is considered first among the 2HD models. The HL-LHC has a high discovery potential for parameter regions with large tan$\beta$. In contrast, the deviation of the Higgs boson coupling with the gauge bosons, g(HZZ) or g(HWW), becomes larger for smaller tan$\beta$, which is favorable for the ILC. The sensitivity of direct searches at the LHC and the ILC250 sensitivity are thus complementary. Heavy Higgs bosons (or SUSY breaking scale $\Lambda$) can be discovered almost up to 1.5–2 TeV by combining the ILC250 and the HL-LHC results, even if the supersymmetric partners are heavy. In the extended models such as the Next-to-Minimal Supersymmetric Standard Model (NMSSM), in which the relation between the neutral and charged Higgs bosons become model-dependent, the large tan$\beta$ region is covered by neutral Higgs boson searches at the HL-LHC and the charged Higgs boson searches at Belle II; the small tan$\beta$ region is covered by the ILC250.

The energy scale of new phenomena can be probed up to $\Lambda$~2 TeV in more general 2HD models not restricted to supersymmetry, through coupling deviations. The Kaluza-Klein (KK) gluon can be probed up to a mass of 10–20 TeV (corresponding to a KK scale of 3–7 TeV).

Furthermore, a physics model behind the discovered new phenomena at Λ_can be identified through the deviation pattern. These sensitivities are determined by the precision of the Higgs couplings, and do not depend highly on the CM energy (250 GeV or 500 GeV). It will be crucial to reduce the systematic uncertainties (coming from experimental uncertainties, and determining the quark mass and $\alpha_s$) through the collaboration between the experimental and theory communities.

## 2.2. Precision measurement of the Higgs boson properties

The total decay width $\Gamma_H$ can be determined with an accuracy of 2.1% by fitting the both results at ILC250 and HL-LHC in the effective field theory framework as mentioned in Section 2.1. This will allow for the search for decays to unknown particles with decay branching ratios down to 0.3%. Detail is discussed in Chapter 4. The CP phase in the coupling between the Higgs boson and fermions can be measured to 3.8 degree precision, which provides an important clue to the origin of the matter-antimatter asymmetry, whether it is baryogenesis or leptogenesis (See Chapter 3). The discovery of CP violation in the Higgs sector will be an important achievement, as it implies that the SM Higgs field (1HD) is not the correct description of nature, and that the Higgs sector must be more complicated (such as general 2HD models or addition of singlet fields).

From the angular distribution of the decay particles, the compositeness of the Higgs boson can be probed up to a scale of 2.2 TeV.

## 2.3. Precision observables in the Standard Model: $M_W$ / $M_t$ / $\sin\theta_{eff}$

At ILC250, the W boson mass ($M_W$) and weak mixing angle ($\sin\theta_{eff}$) can be measured with accuracies of 3 MeV and $3\times10^{-5}$ (relative precision), respectively. Although the top quark mass ($M_t$) cannot be measured directly at the ILC250, the HL-LHC is expected to determine the top quark mass with an accuracy of 0.2–0.3 GeV. The contribution of various systematic uncertainties such as $\Delta M_Z$ and $\Delta\alpha_s$ and a top quark mass precision of 0.3 GeV are roughly equal to check the Standard Model precisely. Thus, from the point of view of precise observables in the SM, the HL-LHC precision of 0.3 GeV is sufficient. Supposing that the current central values for $M_W$, $M_t$, and $\sin\theta_{eff}$ remain fixed, the improved precision from the ILC250 and the HL-LHC will yield a 3–4$\sigma$ deviation from the SM. This will indicate that new physics such as supersymmetry exists around the TeV scale. If an excess is seen at the HL-LHC or if deviations of Higgs couplings are seen at the ILC, it will be crucial to identify the principles behind these anomalies. This will be one of the important achievements expected from the ILC250.

The stability of our vacuum can be computed from the Higgs boson mass ($M_h$) and $M_t$. An upper limit on the energy scale of new physics can be also determined with the assumption that our vacuum is stable. Combining the ILC precision of $\Delta M_h$=14 MeV and the HL-LHC precision of $\Delta M_t$=0.3 GeV will determine that our universe is metastable or that new physics should exist at a scale below $10^{12}$ GeV to make our universe stable, if the central values are the same as the current values. These results are crucial to understand the early universe, including implications about the possibility of leptogenesis, as described in Chapter 3.

2.4. New phenomena can be discovered up to $\Lambda$=2–3 TeV with synergy among the ILC250, HL-LHC and the SuperKEKB. The ILC250 has high sensitivity in the region that cannot be covered with the HL-LHC (heavy higgs boson in the 2HD models with small tan$\beta$ and the electroweak gaugino). The ILC is therefore complementary to the HL-LHC. If an excess is found at the HL-LHC, ILC can play an important role to reveal the physics behind it.

Figure 2 shows a flowchart of overview. If a deviation from the Standard Model prediction is observed in the Higgs coupling, the EW precise measurements or searches, the new energy scale $\Lambda$ for the new phenomena and new principles is determined. It also fixes the technology and the CM energy of the next-generation accelerators, such as Future Circular Colliders (FCC, HE-LHC) and the energy upgrade of the ILC.

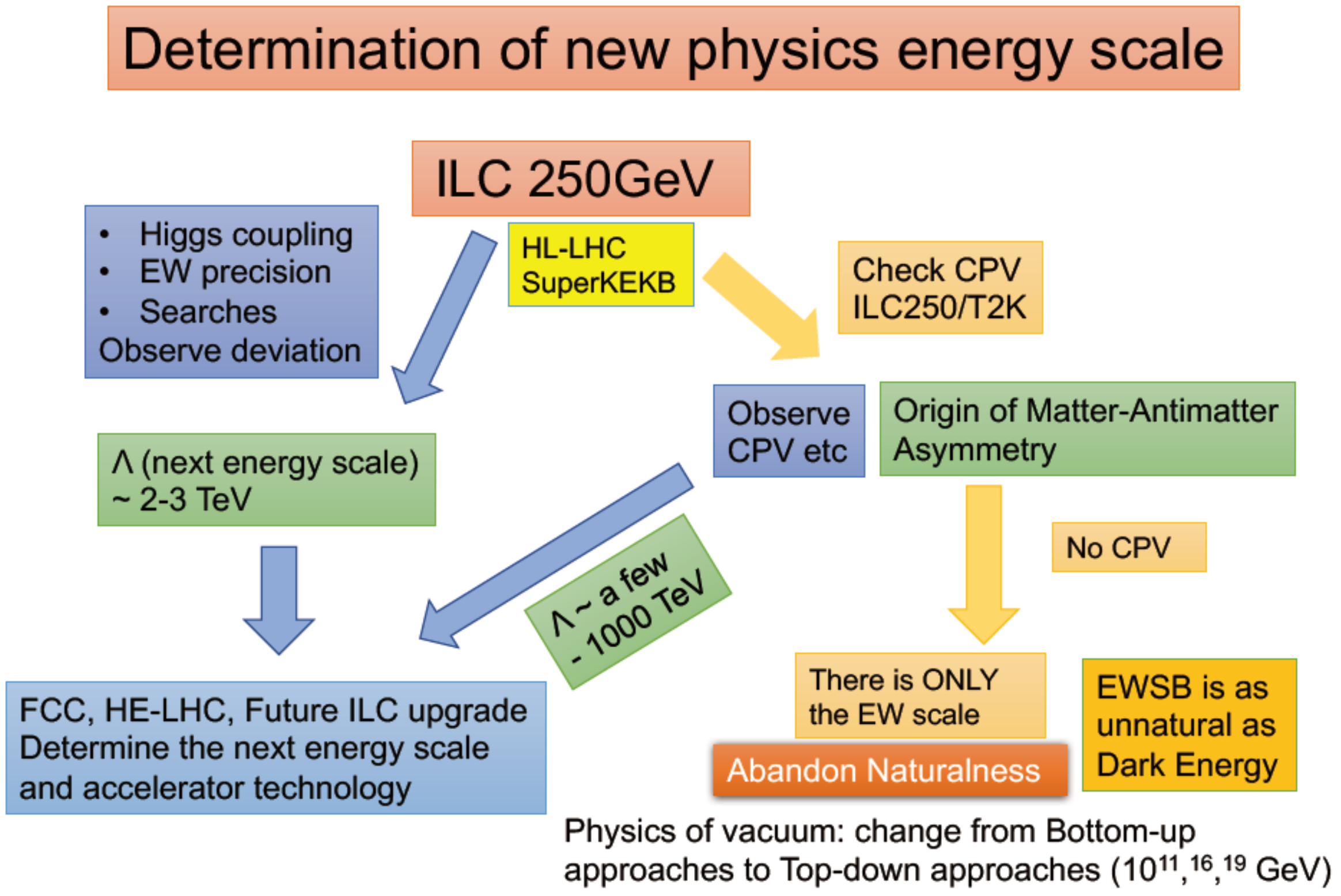

Figure 2: Precision measurements and the energy scale $\Lambda$.

The right side of Figure 2 illustrates the different approach, in which the probing for the origin of the matter-antimatter asymmetry will lead to the next new phenomena (where $\Lambda$ is the new energy scale). As discussed in Chapter 3, it can be determined that the origin of matter is either electroweak baryogenesis ($\Lambda$=10–1000 TeV) or leptogenesis ($\Lambda$<10 TeV) with the observation of CP violation (in Higgs or neutrino sectors), the precise measurement of the Higgs boson, and measurements of gravitational waves in space.

If the results from the ILC250 and other experiments are found to be consistent with the SM, and no new sources of CP violation are found, it will be determined that the energy scale for the physics behind the EWSB mechanism is O(10) times higher than the EWSB scale itself. It will be found to be at least O(10) times "unnatural" , and the electroweak phase transition (EWPT) will be an as unnatural phenomenon as dark energy. The EWPT is related to the vacuum physics, as the same as the inflation and dark energy, whose scales are also not naturally explained. This will lead to a paradigm shift in our research direction, from the traditional bottom-up approach to the top-down approach.

## 3. Electroweak Symmetry Breaking and the Origin of Matter-Antimatter Asymmetry

There are two promising scenarios for the origin of matter-antimatter asymmetry; leptogenesis and electroweak baryogenesis (EWBG). Figure 3 shows a flowchart for approaching these scenarios. EWBG can be probed at the ILC250 in two phases as discussed below.

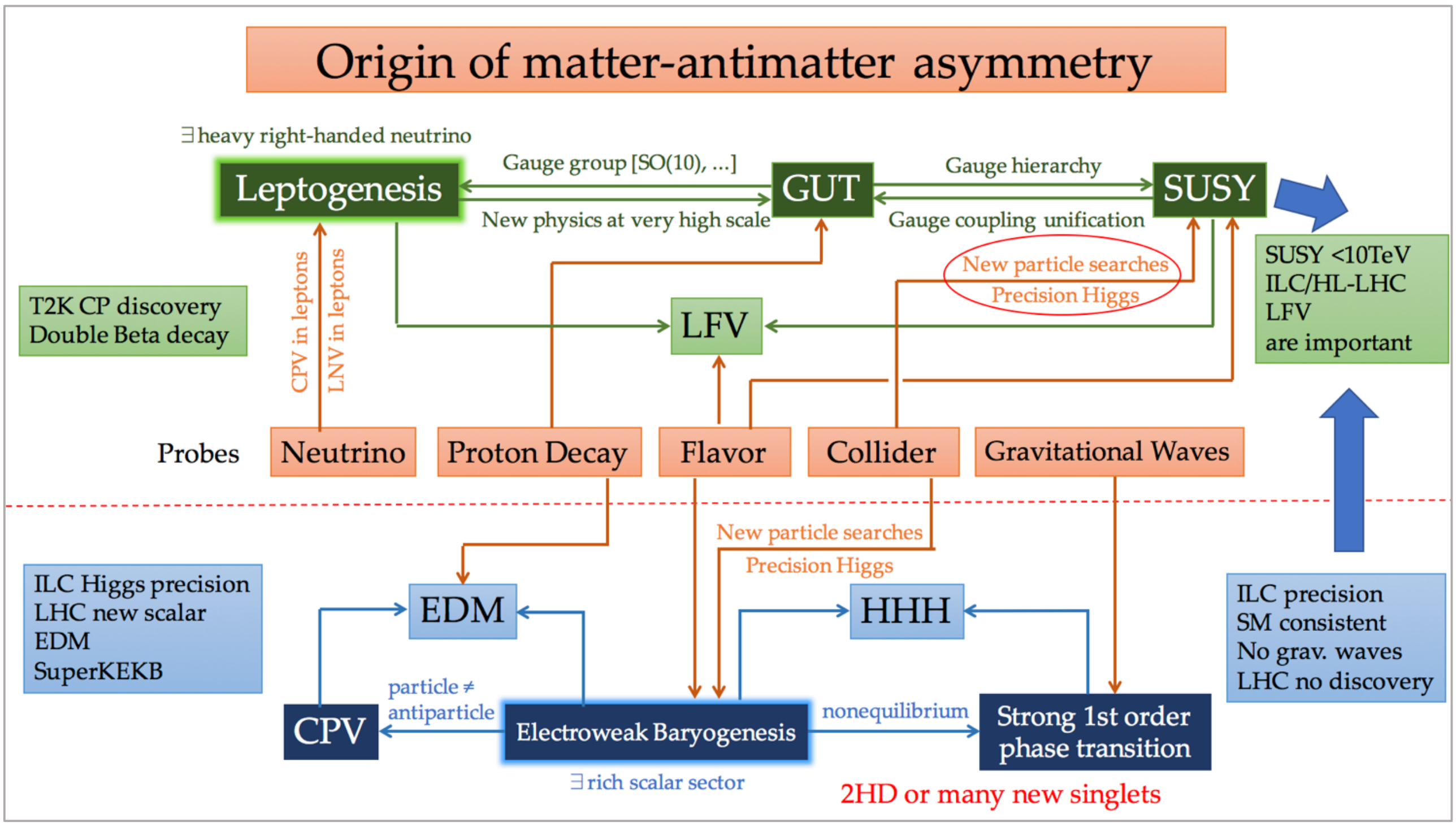

Figure 3: Origin of matter-antimatter asymmetry.

As described in Section 2.2, the phase of the couplings between the Higgs boson and fermions can be measured to 3.8-degree precision. Together with the precise measurements of the HZZ coupling and Yukawa couplings, it can be determined whether or not the Higgs boson is responsible for the origin of matter-antimatter asymmetry. These results can be cross-checked with experiments searching for the electric dipole moment of the neutron and the electron. This is the first step to examine EWBG (left-lower side of Figure 3) .

Furthermore, the electroweak symmetry breaking should be a strong first-order phase transition and in non-equilibrium in order to retain the asymmetry produced by the Higgs sector. For this to occur, it is necessary to introduce additional Higgs fields as in 2HD models or an additional singlet. These new scalar fields result in large deviations (>20%) in the trilinear Higgs coupling and most probably a few-percent deviation in Higgs couplings with gauge bosons as well (right-lower side of Figure.3). The ILC250 can investigate these couplings at sufficient precision. Gravitational waves are also emitted during the EWPT in Strong 1$^{st}$ order transition. They can be detected at satellite probes (such as LISA and DECIGO) which are expected to begin operation around 2040. This is the second step to examine EWBG. The Higgs potential demands that new phenomena must be present in the range up to Λ=a few–1000 TeV in the case of EWBG. The energy of the next energy frontier experiment and its accelerator technology can be also determined in this case (as shown in Fig.2)

The ILC250 can examine EWBG scenarios from many sides. The key to probe EWBG scenarios is the precise measurement of CP violation, the Higgs couplings with the gauge bosons, and the Yukawa couplings. The precisions of these measurements are largely similar between the ILC250 and the ILC500 results. Although the Higgs trilinear coupling (HHH coupling) cannot be measured at the ILC250, the precise measurement of the Higgs boson couplings with the gauge bosons and the gravitational wave probes can be used to elucidate the origin of matter. The crucial test of EWBG can be performed at the ILC250.

If the EWBG scenarios are disfavored at the ILC250, or if CP violation is observed in neutrino sector at the T2K experiment and neutrino-less double-beta decay is discovered, the leptogenesis scenario becomes favorable. This implys the existence of a right-handed neutrino at a very high energy scale as well as grand unification (GUT). The most attractive scenario for GUT is supersymmetry with gauginos and higgsinos under 10 TeV. This scenario can be examined with the search for lepton number violation, the search for proton decays at the HyperKamiokande experiment, and the search for gauginos and higgsinos under 10 TeV at the next hadron collider (FCC, HE-LHC, etc.) or at a higher energy lepton collider. The favorability of the leptogenesis scenario is important input for the discussion of the accelerator technology and the CM energy of the next-generation facility. This is the path labeled "Λ~ a few–1000 TeV" in Figure 2. Since this is an important scenario, it is in the interest of Japan's long-term strategy to construct a linear collider which can easily accommodate the next-generation technology.

## 4. Direct Search for Dark Matter and New Particles Based on Naturalness

Naturalness has played an important role in the history of particle physics. The discovery of the 125 GeV Higgs boson has started to cast some doubt to this idea in the current situation that no new physics is found. The claim is that using supersymmetry to explain a 125 GeV Higgs boson nominally requires fine-tuning on the order of around O(100)–O(1000). However, there are possibilities where squarks become naturally heavy like focus point models. The Higgs boson becomes also naturally heavy in extensions of the MSSM, with additional singlets for example. Before giving up on the idea of naturalness, these possibilities (Higgsino/Wino/Singlet-like ) must be probed. They are also scenarios that provide natural candidates of dark matter (DM). Figure 4 summarizes the candidates of WIMP DM and their searches.

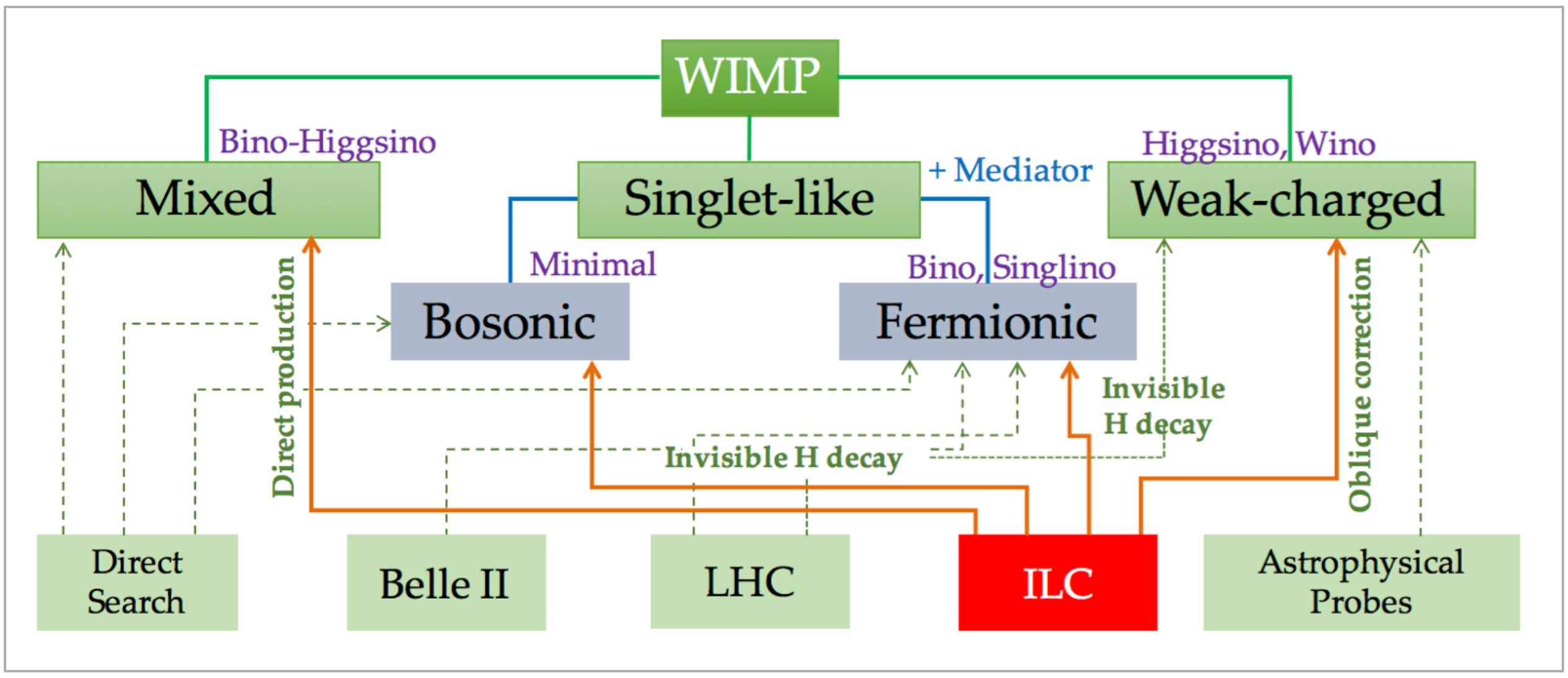

Figure 4: WIMP dark matter candidates and searches.

For the electroweak gauginos and higgsinos, the mass differences from the lightest particle among them (wino, higgsino) are generally small as described in Section 2.1(b). Such compressed spectra are challenging to search for at the LHC. It is also possible that the existence of a singlet particle could suppress the couplings with the gauge bosons. The investigation of these two possibilities are important. Taking together the HL-LHC searches (having a sensitivity for the bino) and the approach of 2.1(b) will make the search strategy complete.

It should be noted that the higgsino mass should be naturally around the Higgs mass; thus the search for the higgsino is particularly important. The ILC250 can probe higgsinos indirectly up to about 200 GeV, corresponding to the test of naturalness of about 10%. This is the higgsino path shown in Figure 4. Since it is challenging to search for higgsinos at the HL-LHC, the ILC searches are indispensable.

In the case of a singlet-like DM as in the middle of Figure 4, there are bosonic and fermionic DM (bino-like or singlino-like). A bino-like DM can be covered by gaugino searches at the HL-LHC. For singlino-like DM or for bosonic DM, the Higgs invisible decays to unknown particles is important. The ILC250 is sensitive to Higgs invisible decays with a branching ratio of 0.3%. This will be a tight constraint for DM candidates lighter than 62 GeV. Such light DM is challenging for underground direct detection experiments because the recoiling energy is small. The ILC will provide coverage for such particles, which makes the search strategy for DM complete.

In the case of a mixed bino and higgsino (Figure 4, left), the strategy depends on its mass. For a heavy mixed bino-higgsino, it can be covered through the gaugino search at the HL-LHC as well as at direct DM detection experiments. If it is light, it will become challenging for the HL-LHC (due to the compressed spectrum) and direct detection experiments (small recoil energy). The ILC250 is able to cover most of the remaining parameter space for these particles by the direct search.

For DM searches based on the "naturalness", the ILC250 will be able to cover regions that cannot be covered by the HL-LHC and direct detection experiments (up to 200 GeV higgsinos and up to 62 GeV singlet-like DM). The ILC250 together with these experiments make the approach of 2.1(b) and the search strategy for WIMP DM complete.

The ILC will play a crucial role in the search for electroweak gauginos and higgsinos and singlet particles motivated by naturalness and dark matter. It plays a complementary role to the HL-LHC and the DM direct detection experiments. Combining these three approaches makes the search strategy complete. The necessary CM energy of the ILC depends on how much fine-tuning one can test for the naturalness.

## 5. Summary

The contributions of each project are summarized in Table 3. As discussed in Chapter 2, the ILC250 will be able to explore the phase space of new physics that cannot be covered by the HL-LHC or the Belle II experiments. It will be able to probe the new phenomena in a robust way. In particular, the ILC250 has an excellent sensitive to the heavy Higgs bosons in 2HD models, which is the 3$^{rd}$ approach as mentioned in Section 2.1(c). The ILC250 has also a good sensitivity to search for dark matter based on naturalness described in Section 2.1(b). Combining three approaches ((a)-(c)) by the ILC250 and the HL-LHC will establish a comprehensive search network, capable of probing the energy scale of new phenomena and new principles (up to $\Lambda$~2–3 TeV). Thus the ILC250 will play an important role.

Table 3: The role of each project.

| | |
|---|---|
| ILC | Higgs & other SM precision measurements; electroweak baryogenesis; 2.1(b): higgsinos, and DM ligher than 62 GeV; 2.1(c): small tan$\beta$. |
| HL-LHC | Higgs couplings; direct search of new phenomena; top quark mass; 2.1(a),(b): bino, wino; 2.1(c): large tan$\beta$. |
| SuperKEKB | Additional CP violation in quark-sector; bottom quark mass; tau LFV (GUT); 2.1(c): large tan$\beta$. |
| T2K, HK | CPV in neutrino-sector; leptogenesis; GUT. |
| LFV | Leptogenesis; right-handed neutrinos; GUT. |
| EDM | Flavor-conserving additional CP violation; electroweak baryogenesis. |
| LISA, DECIGO | First-order phase transition for electroweak baryogenesis: an alternative to the HHH coupling measurement. |
| Underground experiments | DM direct search; 2.1(b): heavy regions. |

The ILC250 will also be able to elucidate the origin of matter. It can perform a crucial test of the electroweak baryogenesis models, and probe the energy scale of new phenomena and new principles (up to $\Lambda\sim$ a few–1000 TeV).

Table 4 summarizes the list of measurements that become challenging by lowering the ILC starting energy to 250 GeV. As far as the precision measurements of the Higgs and other SM observables are considered, the ILC250 operating together with the HL-LHC and the SuperKEKB experiments will be able to play a sufficient role, with precisions not too far from the ILC500.

Table 4: List of measurements that become challenging by making the ILC starting energy 250 GeV.

| Observable | Solutions with synergy |
|---|---|
| Higgs Full Width | From HL-LHC, use custodial symmetry ($K_W/K_Z = 1$) to replace $\Gamma_{HZZ}$ with $\Gamma_{HWW}$ in $\Gamma_{total}=\Gamma_{HWW}/Br(H \rightarrow WW)$<br>→ becomes **comparable to ILC500 precision** |
| Self-coupling HHH (also challenging for ILC500) | Baryon number violation → EWBG or leptogenesis **(T2K, neutrino-less double beta decay).** EWBG covered by **HL-LHC, ILC250, SuperKEKB, LISA.**<br>Although direct measurement of self-coupling is not possible, ILC250 can contribute to examine EWBG through CPV in Higgs sector and the precise measurements of Higgs couplings. |
| Higgs couplings | **HL-LHC** (Top Yukawa coupling)<br>**Lattice** ($m_b$, $m_c$, $\alpha_s$ uncertainty) → **comparable to ILC500**<br>**SuperKEKB** (Lattice examination) |
| Searches | Electroweak gauginos/higgsinos based on naturalness:<br>higgsino (< ~200 GeV); dark matter (<62 GeV). |
| Top mass | HL-LHC (0.2–0.3 GeV) sufficient precision for test of SM; roughly sufficient for vacuum stability; **(if a detailed study of high scale physics becomes necessary, upgrade to 350 GeV)** |

Some of the main merits of the ILC operating at 350 GeV, 500 GeV, or above are

(a) When the energy scale of the new phenomena and new principles is discovered by the combined results of the ILC250 and HL-LHC, this energy scale becomes the next target for an energy upgrade of ILC.

(b) Top quark mass precise measurement: The HL-LHC precision of 0.2–0.3 GeV is sufficient for the test of the SM and the vacuum stability. If the results from the HL-LHC and ILC250 point to physics at very high energy scales such as GUT and the necessity to study the vacuum stability in further detail, then the ILC350 becomes important.

(c) When only the electroweak scale seems to exist (the scenario in Figure 2 (right)), it becomes important to directly study the breaking of the electroweak symmetry and the Higgs potential in detail. In this case, the measurement of the Higgs self-coupling (HHH) becomes important, irrespective to the indirect measurement by gravitational waves. The precise measurement at CM energy of 500 GeV (positive interference) and 1 TeV (negative interference) will be both important.

Future energy upgrade scenarios should be discussed based on the findings of the energy scale of new phenomena and new principles as in point (a) above, or the CM energy will be upgraded, as before, upto 350, 500 GeV or 1TeV based on points (b) and (c).

## Conclusions

The conclusions of this committee are the following four points:

- In order to maximally exploit the potential of the HL-LHC measurements, concurrent running of the ILC250 is crucial.
- LHC has not yet discovered new phenomena beyond the Standard Model. The ILC250 operating as a Higgs Factory will play an indispensable role to fully cover new phenomena up to $\Lambda$~2–3 TeV and uncover the origin of matter-antimatter asymmetry, combing all the results of ILC250, HL-LHC, the SuperKEKB, and other experiments. Synergy is a key.
- Given that a new physics scale is yet to be found, ILC250 is expected to deliver physics outcomes, combined with those at HL-LHC, SuperKEKB and other experiments, that are nearly comparable to those previously estimated for ILC500 in precise examinations of the Higgs boson and the Standard Model.
- The inherent advantage of a linear collider is its energy upgradability. The ILC250 has the potential, through an energy upgrade, to reach the energy scale of the new physics discovered by its own physics program.

Many thanks to T.Tanabe (ICEPP, U.Tokyo), T.Nakada (EPFL) , H.Aihara (U.Tokyo) and S.Komamiya (U.Tokyo) for useful discussions and suggestions to translate the original document into English.